\documentclass{amsart}
\usepackage{mathtools}
\usepackage{amssymb}
\usepackage{amsthm}
\usepackage{amsmath}

\DeclareMathOperator{\KL}{D_{KL}}
\DeclareMathOperator{\Ent}{H}

\begin{document}

\title{Investing is Compression}

\author[Oscar Stiffelman]{Oscar Stiffelman \\ Nand Capital \\
\lowercase{\texttt{ozzie@cs.stanford.edu}} \\
\lowercase{\texttt{imagine@gmail.com}} }

% amsart version %\author{Oscar Stiffelman} %\address{Nand Capital}
%\email{ozzie@cs.stanford.edu} %\email{imagine@gmail.com}

% article version %\author{Oscar Stiffelman \\ % \small Nand Capital
% \small ozzie@cs.stanford.edu \\ % \small imagine@gmail.com}

% \date{\today}

\begin{abstract}
In 1956 John Kelly wrote a paper at Bell Labs describing the
relationship between gambling and information theory. What came to be
known as the Kelly criterion is both an objective and a closed-form
solution to sizing wagers when odds and edge are known. Samuelson
argued it was arbitrary and subjective, and successfully kept it out
of mainstream economics. Luckily it lived on in computer science,
mostly because of Tom Cover's work at Stanford. He showed that it is
the uniquely optimal way to allocate capital given uncertainty: it maximizes long-term wealth,
minimizes the risk of ruin, and is competitively optimal in a
game-theoretic sense, even over the short term.

One of Cover's most surprising contributions to portfolio theory was
the universal portfolio. Related to universal compression in
information theory, it performs asymptotically as well as the best
constant-rebalanced portfolio in hindsight. I borrow a trick from that
algorithm to show that Kelly's objective, even in the general form,
factors the problem into three terms: a money term, an
entropy term, and a divergence term. The only way to maximize growth
is to minimize divergence which measures the difference between our
distribution and the true distribution in bits. Investing is,
fundamentally, a compression problem.

This decomposition also yields new practical results. Because the
money and entropy terms are constant across strategies in a given
backtest, the difference in log growth between two strategies measures
their relative divergence in bits, up to bounded correction terms. I also introduce a winner fraction
heuristic which allocates capital in proportion to each asset's
probability of dominating the candidate set. The growth shortfall of
this heuristic relative to the optimal portfolio is bounded by the
entropy of the winner fraction distribution. To my knowledge, both the
heuristic and the entropy bound are original contributions.

\end{abstract}

\maketitle

\section{Introduction}

In 1956 John Kelly wrote a paper while working at Bell
Labs~\cite{kelly1956} describing the relationship between gambling and
information theory. The paper describes what would eventually be known
as the Kelly criterion. The term Kelly criterion has come to signify
two distinct things: the objective or utility function, as well as the
closed-form solution to the problem when a gambler's odds and edge are
known. Kelly observed that simply maximizing expected wealth would
surely lead to ruin given enough time. What was needed was some kind
of nonlinear utility function to balance risk and reward. The math of
utility functions came from von Neumann and Morgenstern's game
theory. It formalizes a subjective preference or willingness to risk
loss in pursuit of gain. Kelly wanted to avoid the ``introduction of
an arbitrary cost function,'' likening it to the external value that
people attempted to attach to Shannon's information. Kelly realized
that maximizing logarithmic growth rather than simple expectation
simultaneously avoids ruin while maximizing asymptotic wealth. Kelly
wasn't the first to propose this objective. It was originally proposed
by Bernoulli in 1738 as the solution to the Saint Petersburg
paradox \cite{bernoulli1738}. But Kelly was the first to connect it to
the new field of information theory, encouraged by Shannon himself who
assisted with the paper and immediately realized the significance.

It wasn't clear at the time what was so special about this particular
function. It looks a little bit like Shannon's celebrated entropy
equation, but with money replacing the term inside the logarithm. The function
balances potential gains and losses along with their probabilities
through the curvature of the log function. Because log is steeper to
the left than to the right at every point, losses matter more than
gains. And log of zero is negative infinity, so it's not enough to
have a small probability of blowup (ruin) -- it has to be strictly
zero. Another interpretation is that investments should be both
structurally and statistically asymmetric.

Shannon shared the paper with Edward Thorp, who applied it first to
blackjack \cite{thorp1966beat}, then to the stock market through
Princeton Newport Partners, building one of the earliest quantitative investment funds
and achieving one of the most successful track records in investing
history \cite{thorp2017man}. Thorp also conducted extensive
mathematical analysis of the Kelly criterion, extending it to
practical multi-asset and options settings. \cite{maclean2011kelly}

Shannon separately presented these ideas to the economist Paul Samuelson at
MIT. Over the course of several evening meetings \cite{cover1998shannon}, they discussed and
debated the merits of this proposal. Samuelson is said to have had a
``visceral dislike of log utility.'' He argued it was an arbitrary and
subjective utility function, and an unnecessarily conservative one at
that. Over the course of several papers, he defended economics against
this incursion by computer science and a ``fallacy that has been
borrowed into portfolio theory from information theory of the Shannon
type.'' \cite{samuelson1969, samuelson1971, merton1974, samuelson1979} As perhaps the most influential economist at the time, he
successfully kept this idea out of mainstream economics.

Thankfully, this idea lived on in computer science, most
notably because of the work of Tom Cover, the Stanford information
theory professor. We now know because of Cover's work that Shannon's
idea was not a subjective, arbitrary utility function as Samuelson had
argued. It is objectively optimal for three reasons.

\begin{enumerate}
\item It maximizes long-term wealth.

\item It minimizes the risk of ruin.

\item In a game-theoretic sense, it is competitively optimal, even
over the short term. Not just asymptotically as Samuelson
believed. \cite{bell1980, bell1988, algoet1988}
\end{enumerate}

Samuelson seems to have understood the first two points, but he
remained unconvinced, believing there was nothing special about log
that distinguished it from other utility functions. The third reason,
the most important, wasn't known until much later because of Cover's
work at Stanford. Proving optimality required new ideas and techniques
in information theory, including a version of the asymptotic
equipartition property (AEP) for markets to show that it dominates
every other strategy over the long run \cite{algoet1988}, and an
application of Kuhn-Tucker conditions \cite{bell1988} to show that it
is also optimal in a game-theoretic sense over the short term. Cover
noted that it was ``tantalizing that [the log-optimal strategy] arises
as the solution to such dissimilar problems'' and wondered if it was
just a mathematical coincidence or something more fundamental.
\cite{bell1980}  As Cover observed: \begin{quote} ``there is no essential conflict between good short-term
and long-run performance. Both are achieved by maximizing the
conditional expected log return.'' Bell and Cover~\cite{bell1988}
\end{quote}

It might seem hard to reconcile the optimality of Kelly with its
reputation for wild trajectories, extreme volatility, and substantial
drawdowns. Some of this perception comes from miscalibration: when we
overestimate our edge or signal-to-noise ratio, the variance drag from
excess leverage can crush growth. But even with accurate models, Kelly
can follow trajectories that exceed an investor’s risk tolerance. To
understand why this is consistent with its optimality, distinguish the
long-run and short-run claims. In the long run, Kelly dominates every
other strategy over almost every market sequence. The short-run claim is more subtle. It
is a game-theoretic statement: no rival strategy can be
expected to beat Kelly over any horizon. It is a statement about the optimality of the
strategy itself rather than a guarantee about the finite, realized
wealth path. Between the short run and long run, there can be
significant drawdowns. Many turn to ``fractional Kelly'' to dampen the
swings. I argue that this is exactly the right approach, and this
work helps explain why that is justified.

Kelly is both scale invariant and path blind. At every point it chooses the optimal
allocation to maximize growth in a purely forward-looking sense. From
a risk perspective, we are often anchored by the past and we want to
limit drawdowns or broad classes of portfolio trajectories. To limit
drawdowns we need to reduce risk as losses accumulate. This paper
shows that every allocation of capital factors the investing problem
into three terms, including an excess divergence penalty, similar to a
variance drag. Because fractional Kelly is just full Kelly on a
fraction of the capital, it is the only allocation of the at-risk
capital that does not introduce an excess drag on its growth.

One of Cover's most surprising contributions to portfolio theory was
something called a universal portfolio. Related to universal
compression in information theory, it performs asymptotically as well
as the best possible adversary. One that knows the rules but not the
future. Although the math is elegant, the argument is very
abstract. But one of the steps that Cover used in creating universal
portfolios is extremely useful beyond that algorithm. It reveals the
information structure of the investing problem, making it accessible
to the techniques of information theory.

That same technique will be applied here to show that what is so
powerful about the Kelly criterion is that even in the most general
form, it factors the investing problem into three terms: a money term,
an entropy term, and a divergence term. The first two terms don't
depend on our allocation or distribution. The only way to
maximize our compounding growth rate is to minimize the
nonnegative drag or friction from that divergence term which is measured in bits
and only goes to zero when our distribution matches the unknown true
distribution. It is the excess expected code length.
Because minimizing that encoding cost is the only way to maximize growth,
investing is fundamentally a compression problem.

The Kelly criterion was originally presented as a closed-form solution
to games of chance where odds and edge are known. This work revisits
that simple form as well as a generalization known as a horse race
market where there are multiple choices but only one winner, with all
other choices going to zero. The single-winner case is especially
accessible to information-theoretic analysis because exclusivity makes
the growth expression easy to factor. But the main focus of
this work is the general case where, like the stock market, there can be
many nonzero returns. This is the general investing problem where the
challenge is not to identify a single winner but rather an optimal
joint allocation of capital across the available assets. The choices
can include risk-free instruments, leverage, and even shorting. As
with the simple cases, the portfolio represents a probability
distribution by construction. In the general form it is no longer a
simple probability of winning but rather something more abstract. The
math admits multiple interpretations.

\section{Classic Kelly}

The classic Kelly criterion tells us how much to wager on a game of
chance with known odds and predictive edge. Assume that with
probability $p$ we win and our wager is multiplied by return $r$. And
with probability $(1-p)$ we lose the full wager. The question is what
fraction $f$ to wager to maximize wealth. Note that $r$ is the
return multiplier, not the more common $\text{odds} = r - 1$.

If we start with initial wealth $V_0$, then with probability
$p$ we will win and our wealth at the next step will be
\begin{equation}
V_1=V_0 (f r+(1-f))
\end{equation}

and with probability $(1-p)$ we will lose and our wealth will be
\begin{equation}
  V_1 = V_0 (1-f)
\end{equation}

After $n$ steps, our wealth becomes
\begin{equation}
  V_n=V_0 (1-f)^{n_0} (f r + (1-f))^{n_1}
\end{equation}

if we have $n_0$ losses and $n_1$ wins.

Kelly's proposal is that we maximize the expected log wealth, or the
probability-weighted average log return. Note that log is monotonic,
so maximizing log return also maximizes total growth after $n$
steps. Kelly's focus was the expected log growth per step, $g$. Since
each step is multiplicative, the log of the full product is the sum of
the logs, and the average converges to the expectation in the limit.

\begin{equation}
g(f) = \lim_{n \to\infty} \frac{\log \left(\frac{V_n}{V_0}\right)}{n}=p \log (f r+(1-f))+(1-p) \log (1-f)
\end{equation}

We can find the $f^*$ that maximizes $g$ by taking the derivative and
setting it equal to zero

\begin{equation}
\frac{d g}{d f}=\frac{p (r-1)}{f
  r-f+1}-\frac{1-p}{1-f}=0
%
%\frac{\partial g(f)}{\partial f}=\frac{p (r-1)}{f
%  r-f+1}-\frac{1-p}{1-f}=0
\end{equation}

Solving for $f$ we get

\begin{equation}
  f^* = \frac{p r-1}{r-1}
\end{equation}

This is the famous ``edge/odds'' Kelly solution that maximizes growth.
Plugging it back into the growth equation:

\begin{equation}
g^* = p \log (p r) + (1-p) \log \left(\frac{(1-p) r}{r-1}\right)
\end{equation}

If the market maker's probability is $q$ and the odds are fair, then
$q \cdot r = 1$, and $r = \frac{1}{q}$. Substituting into the equation:

\begin{equation}
g^* = p \log \left(\frac{p}{q}\right) + (1-p) \log \left(\frac{1-p}{1-q}\right)
\end{equation}

This is the KL divergence $\KL(P\|Q)$ between our distribution $p$ and the market's
distribution $q$.

\section {Horse Race Market}

Consider a market where you can bet fractionally on mutually exclusive
events and you must fully allocate your capital. This is described as
a ``horse race market'' \cite{cover2006elements} with only one winner each time. In this case we
need a weight vector $W$ rather than a single wager fraction $f$. If there
are $N$ horses, let $w_i$ be the fraction wagered on the $i$th
horse, let $p_i$ be the probability of the $i$th horse
winning, and let $r_i$ be the return or the wealth ratio for
capital wagered on that horse. Kelly's expected log return for this
game becomes:

\begin{equation}
g(W)=\sum _{i=1}^N p_i \log \left(r_i w_i\right)
\end{equation}

In the classic Kelly example, we used calculus to find the optimal
solution and observed the information-theoretic KL divergence in the
solution. This time we will take a more direct information-theoretic
approach to find and characterize the solution. This problem is simple
to analyze because there is only one nonzero term inside the logarithm
(the single winning horse).

Because the log of a product is the sum of logs, we can rewrite this
as

\begin{align}
  g(W) &=\sum_{i=1}^N p_i \log (r_i w_i) \\
       &=\sum_{i=1}^N p_i \log (r_i)+\sum _{i=1}^N p_i \log
         \left(w_i\right) \\
       &= \sum_{i=1}^N p_i \log(r_i)+\sum _{i=1}^N p_i
         \log \left(\frac{p_i w_i}{p_i}\right) \\
       &= \sum_{i=1}^N p_i \log(r_i) + \sum_{i=1}^N p_i \log
         \left(p_i\right) + \sum_{i=1}^N p_i \log
         \left(\frac{w_i}{p_i}\right) \\
       &=\sum_{i=1}^N p_i \log(r_i) - \sum_{i=1}^N p_i \log
         \left(\frac{1}{p_i}\right) - \sum_{i=1}^N p_i
         \log \left(\frac{p_i}{w_i}\right)  \\
       &=\sum_{i=1}^N p_i \log \left(r_i\right) -\Ent(P) - \KL(P\| W)
\end{align}

Notice that the first two terms don't depend on $W$. The only way to
maximize growth is to minimize the third term which is the KL
divergence between the probability distribution and the weight
vector. Because the weight vector sums to one, we can think of it like
a probability distribution. We know from Gibbs' inequality that
$\KL(P\|W)$ is nonnegative and only goes to zero when $P=W$. In other
words, the optimal strategy is to bet in proportion to the true
probability distribution. This is a counterintuitive fact because it
does not depend on the return distribution, just the probability of
winning.

If the payoffs are priced fairly using a market probability $Q$, then as
before $q_i \cdot r_i = 1$ and so $q_i = \frac{1}{r_i}$.

Substituting that back into the above solution, we have

\begin{align}
g(W) &= \sum_{i=1}^{N} p_i \log \left(\frac{1}{q_i}\right) - \Ent(P) -
       \KL(P\|W) \\
     &= \Ent(P,Q) - \Ent(P) - \KL(P\|W) \\
     &= \KL(P\|Q) - \KL(P\|W)
\end{align}

Which follows from the fact that KL divergence is the difference
between the cross entropy and entropy.

The growth rate is the difference between these two KL divergences:
the divergence of the market from the true distribution minus the
divergence of our allocation from the truth. The intuition is that you
can't make money if the market is priced accurately or if you simply
agree with the market. You have to disagree with the market while
being closer to the true distribution.

\section{General Case}

The horse race market entails mutually exclusive events, with only one
nonzero term inside the logarithm at each step. This makes it
straightforward to analyze using standard mathematical
techniques. Consider now the more general case, such as the stock
market, where there can be more than one nonzero term at each
point. This setting accommodates risk-free instruments, leverage, and
shorting. Because the portfolio weight vector is treated as a
probability distribution, it must be nonnegative and sum to one, but
shorting can be handled by introducing mirror instruments representing
collateralized short positions, so that every instrument return remains nonnegative.

In this case, the return at each point in time is the inner product between the weight vector and the
return vector:

\begin{equation} \sum_{i=1}^{N} r_i w_i
\end{equation}

and the log return is given by

\begin{equation} \log \left( \sum_{i=1}^N r_i w_i \right)
\end{equation}

The horse race market expected log return

\begin{equation} \sum_{i=1}^N p_i \log (r_i w_i)
\end{equation}

is a special case of this more general formulation in which the $i$th
term is nonzero with probability $p_i$. The general case is more
difficult to analyze because the sum doesn't have the same clean
factorization.

To analyze this problem, I'm going to use a technique that was first
used by Tom Cover in his universal portfolio theory
\cite{cover1991universal, cover2006elements} proofs. Although Cover
used the technique for a different purpose, it turns out to be
extremely useful for analyzing the general investing problem. It's
essentially a change of representation that reveals an information
structure to the investing problem and makes it more accessible to
techniques and insights from information theory. Cover used this trick
to make it possible to apply the math of universal compression which
is the surprising idea from information theory that you can achieve
the Shannon compression limit asymptotically even without knowing the
probability distribution ahead of time. In the context of investing,
this seems especially paradoxical, but the math is nevertheless sound
and elegant. Rather than using it to design an algorithm as Cover did,
I'm simply borrowing the technique for the purpose of analysis.

We traditionally think of the multi-period investing problem as a
product of scalars. We can write the return after $n$ periods of some
multi-period investment process as the product of the individual
returns:

\begin{equation} R_n = \prod_{t=1}^n r_t
\end{equation}

where $R_n$ indicates the total growth of the multiplicative process
after $n$ steps.

Assuming we are compounding and rebalancing a fixed-weight portfolio
at each point in time, we can write this full process, where $w_i$ is
the portfolio weight assigned to the $i$th asset, and $r_{i,t}$ is the
return of the $i$th asset at time $t$.

\begin{equation} R_n = \prod_{t=1}^n \sum_{i=1}^m w_i r_{i,t}
\end{equation}

Note that the product of scalars is now seen as a product of sums (dot
products).

Cover's brilliant insight was that we can rewrite the product of sums
as a sum of products. This might seem like a trivially algebraic
change. But as will become clear momentarily, instead of compounding
weighted returns per period, we can think of this as a weighted
average of an exponential number of multi-period returns.

Assume for now that there are only two assets in the system $\{a,
b\}$. The return after one period is

\begin{equation} w_a r_{a,1}+w_b r_{b,1}
\end{equation}

After two periods the return is

\begin{equation} \left(w_a r_{a,1}+w_b r_{b,1}\right) \left(w_a
r_{a,2}+w_b r_{b,2}\right)
\end{equation}

Which we can expand as
\begin{equation} w_a^2 r_{a,1} r_{a,2} + w_b^2 r_{b,1} r_{b,2} + w_a
w_b r_{a,1} r_{b,2} + w_b w_a r_{b,1} r_{a,2}
\end{equation}

We have changed the product of sums into a sum of products. The number
of terms grows exponentially with the number of periods. With every
additional period, we double the number of terms in the two-asset
case.

The symbols $\{a,b\}$ can be combined into $2^n$ sequences of length
$n$. Each such sequence corresponds to a term in the expanded $n$-period sum of products, indicating whether the return at time $t$
should be $a$'s or $b$'s return at that time.

If $n_a$ and $n_b$ indicate the number of times each symbol occurs in a
given sequence, the weight for the sequence is

\begin{equation} w_a^{n_a}w_b^{n_b}
\end{equation}

The returns are more complex because they change over time. For now,
think of them as an additional product that goes with each term.

What we now have is an exponential number of sequences that describe
horse-race-style (mutually exclusive) investments on only one symbol
per period. It's straightforward to show that the weights
$w_a^{n_a}w_b^{n_b}$ across these sequences add up to one. Effectively
we have turned the multi-period compounding and rebalancing process
into a single (non-rebalancing) weighted investment across an
exponential number of multi-period investments. Imagine pre-allocating
capital to an exponential number of processes and compounding serially
within each siloed process.

Let $\mathcal{T}_n(P)$ be the set of sequences of length $n$ with
empirical frequencies given by

\begin{align} P &= \{p_1, \dots, p_m\} \\ \sum_{i=1}^m p_i &= 1 \\ n
p_i &\in \mathbb{N}
\end{align}

Each $\mathcal{T}_n(P)$ defines a type class of sequences. It is the
set of all permutations of sequences of length $n$ with symbol
frequencies matching $P$. This lets us write the total growth equation
as a sum over all type classes

\begin{equation} R_n(W) = \sum_P \prod_{i=1}^m w_i^{n p_i} \sum_{s \in
\mathcal{T}_n(P)} r_s
\end{equation}

where $r_s$ is the product of returns obtained by following sequence
$s$, taking the return of whichever asset the sequence selects at each
time period. This is the same exponential set of sequences as before, just grouped
by type and with the common sequence weight $\prod_{i=1}^m w_i^{n
p_i}$ factored out per class.

As stated earlier, the weights sum to one across the full set of
sequences. Each sequence weight represents a budget allocation to that
particular multi-period compounding process that bets on one specific
symbol per time period. A particular type class represents all
permutations of a given sequence. Because all permutations in the
class have the same weight, it's not possible to wager more on a
particular sequence than any other sequence in the set. The larger the
set (higher the entropy), the less specific the wager. Cover used this
fact to show that we can asymptotically achieve the same growth as an
ideal adversary, one that knows the ``rules'' (the true odds) but not
the future.  The key insight is that even such an adversary is
similarly restricted and can bet with high concentration only if the
set is small (low entropy). By using the entropy of the sequence to
size the wager, one can wager proportionally to the unknown optimal
bet with only a polynomial penalty to make the wagers sum to one (the
cost of universality). From a traditional finance perspective, this is
a truly surprising result. The idea of optimal investing without any
informational advantage seems paradoxical, but it's really a
demonstration of how powerful it is to apply information theory to the
investing problem. Most people who are aware of Cover's universal
portfolio know it as a beautifully abstract and elegant yet likely
impractical algorithm due to transaction costs. But as I'll
demonstrate, this isn't just a scaffold for making universal
portfolios. Cover's idea to rewrite the traditional product-of-sums
investing problem as a sum of products reveals the information
structure of the general investing problem. It's not just useful for
Cover's very abstract idea -- it's an essential tool that I argue
makes the investing problem less opaque and more accessible to the
techniques of information theory.

We will now demonstrate how the weight, which sums to one across all
sequences, concentrates around the particular type class $P=W$, the set of
all permutations of sequences with frequencies matching the portfolio
weights. We first establish this using combinatorics and Stirling's
approximation. Because this approximation is only asymptotically
valid, we then refine the argument using KL divergence to show how the
total weight assigned to non-dominant sets declines exponentially with
both divergence and time. Once this type class concentration is
established we will use it to simplify the growth function to reveal
the information-theoretic terms in the investment process.

The size of $\mathcal{T}_n(W)$ is the number of permutations of
sequences of length $n$ and with $n_i = w_i n$. The size of this set
is given by the multinomial

\begin{equation} \binom{n}{n_1, n_2, n_3, \dots, n_m} = \frac{n!}{n_1!
n_2!  n_3!\dots n_m!} = \frac{n!}{\prod_{i=1}^{m}n_i!}
\end{equation}

For the $\mathcal{T}_n(W)$ type class, this is

\begin{equation} \frac{n!}{\prod_{i=1}^{m}(n w_i)!}
\end{equation}

Using Stirling's approximation

\begin{equation} \log(n!) = n \log(n) -n + \mathcal{O}(\log(n))
\end{equation}

Ignoring the sublinear term (this approximation will be revisited
below), we can rewrite the cardinality, using $\mathcal{T}_n(W)$ to
denote both the cardinality and the set where it's unambiguous:

\begin{align} \log(\mathcal{T}_n(W)) &= \log(n!) -
\log\left(\prod_{i=1}^m (n w_i) !\right) \\ &= n \log(n) - n -
\sum_{i=1}^m(n w_i \log(n w_i) - n w_i) \\ &= n \log(n) - n -
\sum_{i=1}^m n w_i \log(n) - \sum_{i=1}^m n w_i \log(w_i) +
\sum_{i=1}^m n w_i \\ &= n \log(n) - n - n \log(n) \sum_{i=1}^m w_i -
n \sum_{i=1}^m w_i \log(w_i) + n \sum_{i=1}^m w_i \\ &= -n
\sum_{i=1}^m w_i \log(w_i) \\ &= n \sum_{i=1}^m w_i
\log\left(\frac{1}{w_i}\right) \\ &= n \Ent(W)
\end{align}

The cardinality is therefore

\begin{equation} \mathcal{T}_n(W) = e^{n \Ent(W)}
\end{equation}

The weight assigned to a sequence $s$ is denoted by
$\mathbf{w}_s$. This is the product of the portfolio weights for each
period corresponding to the selected symbol. Because all sequences
within the type class $ \mathcal{T}_n(W)$ have the same symbol
frequencies (they are permutations of each other), they share the same
sequence weight. The sequence weight $\mathbf{w}_s$ for each sequence
$s \in \mathcal{T}_n(W)$ is

\begin{equation} \mathbf{w}_s = \prod_{i=1}^m w_i^{n w_i}
\end{equation}

Applying logs, this becomes

\begin{align} \log(\mathbf{w}_s) &= \log\left(\prod_{i=1}^m w_i^{n
w_i}\right) \\ &= \sum_{i=1}^m n w_i \log(w_i) \\ &= -\sum_{i=1}^m n
w_i \log \left(\frac{1}{w_i}\right) \\ &= -n \Ent(W)
\end{align}

The weight for each sequence is therefore

\begin{equation} \mathbf{w}_s = e^{- n \Ent(W)}
\end{equation}

At least for large $n$ (ignoring subexponential factors, with precise refinement later), the sum of this
per-sequence weight across all of the sequences in the set is
approximately

\begin{equation} \sum_{s \in \mathcal{T}_n(W)} \mathbf{w}_s = e^{-n
\Ent(W)} e^{n \Ent(W)} = 1
\end{equation}

This indicates that the weight is concentrated in the set of
sequences from the type class P = W. Strictly, the mass sits in a
neighborhood of type classes around W rather than in that single
class, and the concentration holds in logarithmic rate terms, where
the polynomial factors hidden by Stirling's approximation vanish after
dividing by n. The KL divergence argument below is what sizes that
neighborhood: it shows the weight on any other class falls off
exponentially with its divergence from W.

This concentration argument relied on Stirling's
approximation. Although this is asymptotically accurate and it
identifies the dominant type class, it overstates the finite log
cardinality.

The intuition is that Shannon's entropy, which characterizes an
infinite stochastic source, is greater than the complexity of a finite
object with the same frequency distribution. The finite case is
formally in the domain of Kolmogorov complexity theory
\cite{li2019kolmogorov}, but from a coding perspective, the reason is
because knowing the full distribution and the number of symbols
remaining reduces the uncertainty.

We can use KL divergence to more precisely bound the weight mass
assigned to type classes that are different from $W$. The weight for
each sequence in type class $P$ is $\prod_{i}w_i^{n p_i}$ and the cardinality is
approximately $|\mathcal{T}_n(P)| = e^{n \Ent(P)}$. The log of the total density
assigned to the set of sequences in the class is

\begin{align} \log\left(\prod_{i}w_i^{n p_i} e^{n \Ent(P)} \right) &= \sum(n p_i \log\left(w_i\right)) + n \Ent(P) \\&= -n \Ent(P,W) + n \Ent(P) \\
&= -n \KL(P\|W)
\end{align}

The total weight concentrated on sequences in that set declines
exponentially with the divergence from the portfolio weight vector $W$
and with the number of periods:

\begin{equation} e^{-n \KL(P\|W)}
\end{equation}

In the original expression

\begin{equation} R_n(W) = \sum_P \prod_{i=1}^m w_i^{n p_i} \sum_{s \in
\mathcal{T}_n(P)} r_s
\end{equation}

Each type class $W'$ contributes a nonnegative term

\begin{align} \label{dominantclass}
  e^{-n \Ent(W', W)} \sum_{s \in \mathcal{T}_n(W')} r_s &=
  e^{-n \KL(W'\|W)} e^{-n \Ent(W')} \sum_{s \in \mathcal{T}_n(W')} r_s
\end{align}

which is the average return for the $W'$ type class with an
exponential divergence penalty. Note that the dominant type class does
not have any penalty because the divergence is zero. For a
non-dominant type class to affect the leading growth rate, its return has to grow faster
than the exponential divergence penalty.

Now assume that \(W^*\) is the optimal portfolio weight vector, the portfolio that maximizes the expected log return. Enumerate the type classes in decreasing order of average return. Matching the portfolio weights to the first class gives it no exponential divergence penalty, while no weighted average of the class returns can exceed the largest class average. Thus this class determines the optimal portfolio’s asymptotic growth rate. To leading exponential order, we can therefore write

\begin{align}
R_n(W^*)
&=\sum_P\prod_{i=1}^m (w_i^*)^{np_i}
  \sum_{s\in\mathcal T_n(P)}r_s\\
&\approx e^{-n \Ent(W^*)}
  \sum_{s\in\mathcal T_n(W^*)}r_s. \label{optimalclass}
\end{align}

Here the approximation preserves the limiting per-period log growth.

If we invest using any other weight vector $W \neq W^*$, it will
achieve growth that is no worse than the divergence penalized
growth for $W^*$. Again, the $W^*$ mass and return are both concentrated
around its dominant type class. For $W \neq W^*$, the contribution from
any other type class is nonnegative, so this is a true lower bound.

\begin{align} R_n(W) &= \sum_P \prod_{i=1}^m w_i^{n p_i} \sum_{s \in
                       \mathcal{T}_n(P)} r_s \\
  &\ge \prod_{i=1}^{m} w_i^{n w_i^*} \sum_{s \in
    \mathcal{T}_n(W^*)} r_s \\
  &= e^{-n \KL(W^*\|W)} e^{-n \Ent(W^*)} \sum_{s \in \mathcal{T}_n(W^*)}
    r_s \\
  \log(R_n(W)) &\ge  \log \left( \sum_{s \in
\mathcal{T}_n(W^*)} r_s \right) -n \Ent(W^*) - n \KL(W^*\|W)
\end{align}

Defining $\mathcal{\bar{R}}_{\mathcal{T}(W)}$ as the per-period
geometric growth rate
\begin{equation}
  \mathcal{\bar{R}}_{\mathcal{T}(W)}^n = \sum_{s \in \mathcal{T}_n(W)} r_s
\end{equation}

The expected log return $g(W)$ is therefore

\begin{equation} \label{generalfactor} g(W) = \lim_{n \to \infty}
\frac{\log(R_n(W))}{n} \ge \log \left( \mathcal{\bar{R}}_{\mathcal{T}(W^*)}  \right) - \Ent(W^*) - \KL(W^*\|W)
\end{equation}

The general case is more complicated than the classic Kelly or the
horse race market problem. But we once again see that Kelly's
objective has factored the investing problem into the same three-term
decomposition: a money term (payoff / rules of the game), an entropy
term (uncertainty) and a divergence term. As before, the first two
terms don't depend on our portfolio allocation. The only way to
maximize the compounding growth rate is to minimize the drag or
friction from the divergence which measures, in bits,\footnote{Throughout, log denotes the natural
  logarithm. Results are stated in bits with the understanding that base conversion is a scaling constant.} the difference
between our distribution and the unknown true distribution. Therefore
investing is, fundamentally, a compression problem.

\section{Practical Applications}

\subsection{Backtesting}

One of the challenges when backtesting and A/B testing strategies is
that market data is notoriously nonstationary. Even the log objective
is sensitive to extreme and atypical historical events.

This makes the average difficult to interpret in a statistical
sense when comparing strategies. But recall that the first two terms
in equation \ref{generalfactor} (money term and entropy term) do not
depend on the allocation, so they are constant across strategies in a
given backtest.

\begin{align}
  g(W_A) &=
           \log \left( \mathcal{\bar{R}}_{\mathcal{T}(W^*)}  \right)
           - \Ent(W^*)
           - \KL(W^*\|W_A) + \delta(W_A) \\
  g(W_B) &=
           \log \left( \mathcal{\bar{R}}_{\mathcal{T}(W^*)}  \right)
           - \Ent(W^*)
           - \KL(W^*\|W_B) + \delta(W_B) \\
g(W_B) - g(W_A) &= \KL(W^*\|W_A) - \KL(W^*\|W_B) + \delta(W_B) - \delta(W_A)
\end{align}

Here $\delta(W)$ is the correction to the retained-class lower bound. That lower bound and the optimality of $W^*$ imply $0 \leq \delta(W) \leq \KL(W^*\|W)$.

Thus the difference in log growth measures the reduction in divergence
up to the difference in these corrections. When that correction
difference is negligible for the comparison, the original divergence
interpretation holds approximately. Closeness to \(W^*\) in KL
divergence bounds the absolute approximation error.

\subsection{MDL, Market Prediction, and Strategy Performance}

Markets are high-dimensional, sparse, and nonstationary. This makes
it very difficult to separate signal from noise. It is especially
challenging for techniques based on statistical learning theory
\cite{vapnik1998} which rely on resampling from a stationary
stochastic source to detect overfitting and generalization. Although
statistical learning theory is the dominant machine learning paradigm,
it's not the only one. Information theory offers alternative learning
paradigms, including the MDL principle \cite{grunwald2007mdl,
li2019kolmogorov}; see \cite{stiffelman2014wrongmodeldata} for a
discussion of prediction, compression, and the relationship between
models and data. The key insight is that randomness is incompressible,
so generalization is about finding a parsimonious, compressive
description.  The relationship between strategy performance and
prediction performance is more direct when both are treated as
information-theoretic questions. In other words, it's easier to draw a
clear line between signal improvements and strategy improvements when
both are measured in bits.

\subsection{Winner Fraction and VC-style Investing}
Kelly's objective is concave. This means that there is a unique
solution, and if we can model the joint return distribution, then we
can solve the general case efficiently using convex optimization techniques.

But it's often not feasible to accurately model the joint
distribution. Consider the problem of venture-capital-style
investing. The investor has to decide how much capital to allocate,
often years before any real clarity on future cash flow.

We saw from equation \ref{optimalclass} that the optimal portfolio
concentrates around a particular type class. Finding that optimal type
class is equivalent to solving the convex optimization problem which
is not feasible without a precise joint probability distribution of
future returns. Even without the full return distribution, the
structure of this decomposition motivates a useful heuristic:
winner fraction investing.

The principle is to invest in proportion to the fraction of universes
where a given asset dominates the candidate set. In other words, the
probability that a given startup ``wins'' relative to the rest of the
contemplated startups. Although this is still nontrivial to predict, it
is far less complicated than the full joint distribution -- it does
not require modeling the actual returns, just the probability of winning.

It will now be shown that this heuristic is approximately
optimal. The lower the entropy, or the more extreme the ``wins,'' the
better the approximation.

Let $W^*$ be the growth-optimal portfolio, and let $W^\prime$ be the vector
of winner probabilities, in other words the probability that the $i$th
startup dominates the candidate set.

\begin{equation}
\frac{R_n(W^*)}{ R_n(W^\prime)} \le \frac{e^{-n \Ent(W^*)} \sum_{s \in
  \mathcal{T}_n(W^*)} r_s}{e^{-n \Ent(W^\prime)} \sum_{s \in \mathcal{T}_n(W^\prime)}
  r_s}
\end{equation}

The numerator and denominator represent averages over their
respective type classes. For the numerator, we rely on the fact that
$e^{-n \Ent(W^*)}$ is exact to leading exponential order, as shown in equation \ref{optimalclass} and $e^{-n \Ent(W')}$ is a true lower
    bound as described in equation \ref{dominantclass}, choosing the dominant
    type class for $W^\prime$.

Recall that each sequence is an exclusive bet on only one asset
per time period. A type class represents all sequence permutations
with a given frequency matching the portfolio weight
vector. The largest return in the winner fraction set is constructed
by stringing together the dominant asset per period. This sequence
must be in the winner fraction set. Let $r_\text{max}$ be the return for that
sequence. By construction, it must also be
at least as large as every sequence in the optimal set
$\mathcal{T}_n(W^*)$. Therefore

\begin{equation}
\forall s \in \mathcal{T}_n(W^*) \colon \quad \frac{r_s}{r_\text{max}} \leq 1
\end{equation}

\begin{align}
\frac{R_n(W^*)}{ R_n(W^\prime)} &\leq \frac{e^{-n \Ent(W^*)} \sum_{s \in
                                  \mathcal{T}_n(W^*)} r_s}{e^{-n \Ent(W^\prime)} r_\text{max}}
\\ & \leq \frac{\frac{1}{|\mathcal{T}_n(W^*)|} \sum_{s \in
\mathcal{T}_n(W^*)} \frac{r_s}{r_\text{max}}}{e^{-n \Ent(W^\prime)}} \\ &
                                                                       \leq
                                                                       e^{n
                                                                       \Ent(W^\prime)}
\end{align}

This shows the ratio of the optimal to the heuristic solution.

Taking limits to find expectation, we see
\begin{equation}
  g(W^*) - g(W^\prime) \leq \lim_{n \to \infty} \frac{\log(e^{n \Ent(W^\prime)})}{n}
  = \Ent(W^\prime)
\end{equation}

The difference in growth between the growth-optimal portfolio and the
winner fraction portfolio
is bounded by the entropy of the winner fraction distribution. That
entropy is itself bounded by the log cardinality of the candidate set, so this
heuristic is especially useful when the candidate set is small or
when the return distribution is extreme. To my knowledge, this
heuristic and proof of the entropy bound are both original.

\section{Conclusion}

The Kelly criterion describes how to allocate capital in simple games
of chance with known odds and edge. It is a closed-form, optimal solution to
Kelly's log wealth objective. But it turns out that Kelly's
objective is itself optimal among all objectives. And this fact
reframes the investing problem itself.

In finance, we often think of risk and reward as two different axes:
for a given amount of risk, we maximize return. This goes back to the
1950s and Markowitz's modern portfolio theory (MPT) \cite{markowitz1952}. A concern with
MPT is that markets are heavy-tailed and so diversification (averaging)
may not reduce risk (variance) \cite{taleb2020}. This is fundamentally a statistical
concern. But even if markets were Gaussian, it would still be
suboptimal because Kelly's objective is uniquely optimal. To justify
that, we have to turn from statistics to information theory.

Instead of thinking of risk and reward as two different axes, Kelly
proposed maximizing a single objective, the logarithm of wealth over
uncertainty. In other words the expectation or the
probability-weighted average log return. Samuelson argued against this idea,
successfully keeping it out of mainstream economics.

We now know because of Tom Cover's work at Stanford that this was not
a subjective, arbitrary utility function as Samuelson believed. It is
the objectively optimal way to invest. It is competitively optimal in
a game-theoretic sense, assuming neither party has an information
advantage.

With the help of a transformation from Cover's universal portfolio
algorithm, we now understand what is so powerful about Kelly's
objective. It reveals the information structure of the
investing problem, factoring it into three terms: a money term
(payoff / rules of the game), an entropy term (uncertainty) and a
divergence term. The only way to maximize compounding growth rate is
to minimize the friction from that divergence term. Fundamentally,
investing is a compression problem.

\bibliographystyle{amsplain}
\bibliography{bibliography}

\end{document}